\documentclass[twocolumn]{revtex4-2} 

\usepackage{amsmath}
\usepackage{amsfonts}
\usepackage{graphicx}
\usepackage{float} 
\usepackage{subfigure} 
\usepackage[colorlinks, linkcolor=blue, anchorcolor=blue, citecolor=blue, urlcolor=blue]{hyperref}
\usepackage{lipsum}
\usepackage{orcidlink}

\begin{document}

\title{Non-Hermitian gravitational effects on Bose-Einstein condensate}

\author{Tie-Fu Zhang\orcidlink{0009-0000-0706-1699}$^{1,2}$}
\thanks{These authors contributed equally to this work.}

\author{Chengxi Li\orcidlink{0009-0007-3715-5574}$^{1,2}$}
\thanks{These authors contributed equally to this work.}

\author{Yitong Pei\orcidlink{0000-0002-3385-5940}$^{3,4}$}
\thanks{Corresponding author: Yitong Pei}
\email{yitongpei@foxmail.com}

\author{Kai Liu$^{1,2}$}

\author{Wu-Ming Liu\orcidlink{0000-0002-1179-2061}$^{1,2,5}$}
\email{wmliu@iphy.ac.cn}

\affiliation{$^1$Beijing National Laboratory for Condensed Matter Physics, Institute of Physics, Chinese Academy of Sciences, Beijing 100190, China \\$^2$School of Physical Sciences, University of Chinese Academy of Sciences, Beijing 100049, China \\$^3$School of Mathematics, Nanjing University of Aeronautics and Astronautics, Nanjing 211106, China \\$^4$Key Laboratory of Mathematical Modelling and High Performance, Computing of Air Vehicles (NUAA), MIIT, Nanjing 211106, China \\$^5$Songshan Lake Materials Laboratory, Dongguan 523808, China }

\begin{abstract}
	We investigated the impact of Non-Hermitian gravitational potentials on the spatial distribution of Bose-Einstein condensate (BEC) wave functions. Through numerical solutions of the Gross-Pitaevskii (GP) equation, we observed that the imaginary component of Non-Hermitian gravitational potentials affects the spatial periodicity of the BEC wave function phase, resulting in spatial displacement of the wave function's peak. By formulating equations describing the momentum of the BEC wave function with respect to Non-Hermitian gravitational potential parameters and solving and analyzing them under specific conditions, we provided a reasoned interpretation of the numerical results. Our findings contribute to exploring the physical essence of Non-Hermitian gravitational potentials and their impact on BEC, offering theoretical guidance for potential related experiments. 
\end{abstract}


\pacs{pacs 1, pacs 2}

\maketitle



\emph{Introduction.}---%
The emergence of Non-Hermitian physics has shattered the constraints of traditional quantum mechanics, expanding our understanding of general systems \cite{nh-01-ashida, nh-02-bender, nh-03-Bender_2007, nh-04-cao}. In recent years, Non-Hermitian physics has garnered significant attention across various domains of physics, achieving remarkable progress \cite{nh-new-science-01, nh-new-science-02, nh-new-science-03, nh-new-science-04, nh-new-nature-01, nh-new-nature-02, nh-new-nature-03, nh-new-nature-04, nh-new-prl-01, nh-new-prl-02, nh-new-prl-03, nh-new-prl-04, nh-new-prl-05, nh-new-prl-06, nh-new-np-01, nh-new-np-02, nh-new-np-03, nh-new-prx-01, nh-new-prx-02, nh-new-prx-03}. 
Concurrently, advancements in space technology have facilitated an increasing number of Bose-Einstein condensate (BEC) experiments in the microgravity environment of space stations \cite{mg-new-01, mg-new-02, mg-new-03, mg-new-04, mg-new-05, mg-new-06}. This raises a question: how does gravity influence the evolution of BEC? Furthermore, what novel physical phenomena arise when Non-Hermitian gravitational potentials act on BEC?

In this Letter, we explore the impact of Non-Hermitian gravitational potentials on BEC through the Gross-Pitaevskii (GP) equation \cite{gpe-01, gpe-02, gpe-03, gpe-04, gpe-05}. Utilizing numerical solutions of the GP equation, we compare the variations in the phase distribution of BEC wave functions under different parameters of Non-Hermitian gravitational potentials. The results indicate that the imaginary part of the Non-Hermitian gravitational potential significantly influences the periodic changes in the phase within the main peak of the BEC wave function, leading to spatial displacement of the wave function's peak. The real part of the Non-Hermitian gravitational potential alters the phase distribution outside the main peak of the BEC wave function. To delve into the physical essence of these findings, we formulate equations describing the momentum of the BEC wave function and provide explanatory insights through analysis.

\begin{figure}
	\centering
	\includegraphics[width=0.42 \textwidth]{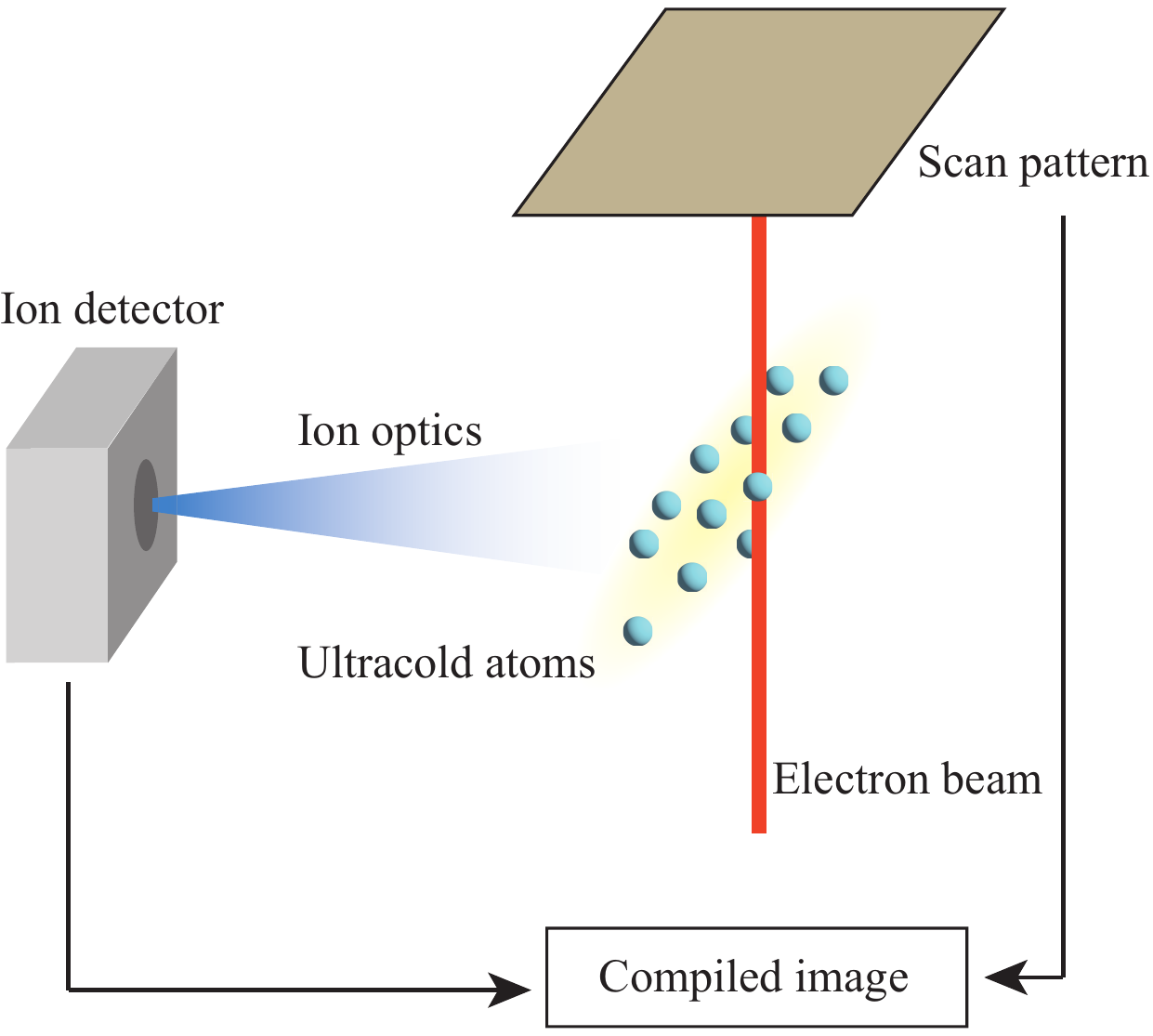}
	\caption{The atomic ensemble is prepared in an optical dipole trap. An electron beam with variable beam current and diameter is scanned across the cloud. Electron impact ionization produces ions, which are guided with an ion optical system towards a channeltron detector. The ion signal together with the scan pattern is used to compile the image.}
	\label{fig-experiment}
\end{figure}

\emph{Simulating Non-Hermitian gravity: experimental setup and control parameters.}---%
We employed electron beam technology \cite{exp-07-lewandowski2012dissipative, exp-01, exp-02, exp-03, exp-04, exp-05, exp-06} to provide experimental simulations of Non-Hermitian gravity (Fig. \ref{fig-experiment}). 
The current density of the electron beam was modeled using a Gaussian beam profile, which can be described by a Gaussian function. 
By adjusting the parameters of the electron beam, such as the current, waist width, and center position, we were able to precisely control the interaction strength and position between the electron beam and the BEC system. 
The BEC system consisted of approximately 
$10^{5}$ atoms confined by an anisotropic harmonic trap. 
The trap had distinct frequencies, $\omega_1 = 2\pi\times 13Hz$ along the $x_1$-axis and $\omega_2 = 2\pi\times 170Hz$ along the $x_2$-axis \cite{exp-07-lewandowski2012dissipative}.
The spatial distribution of the electron beam was described by a Gaussian function, where the waist width $\omega$ could be determined based on the half-width at half-maximum (FWHM). 
By adjusting the waist width and center position of the electron beam, we were able to match the width of the complex potential field with the characteristics of the electron beam. 

The key in the experiment was to introduce an imaginary potential field to simulate the interaction between the electron beam and the BEC system. 
The width of this imaginary potential field was determined by the Gaussian profile of the electron beam, while the interaction strength was determined by the current of the electron beam and the total cross-section of $^{87}\mathrm{Rb}$ atoms. 
By precisely controlling the experimental parameters, we could simulate and study the Bose system under Non-Hermitian gravity, and compare the experimental results with theoretical calculations.

\begin{figure}
	\centering
	\includegraphics[width = 0.45 \textwidth]{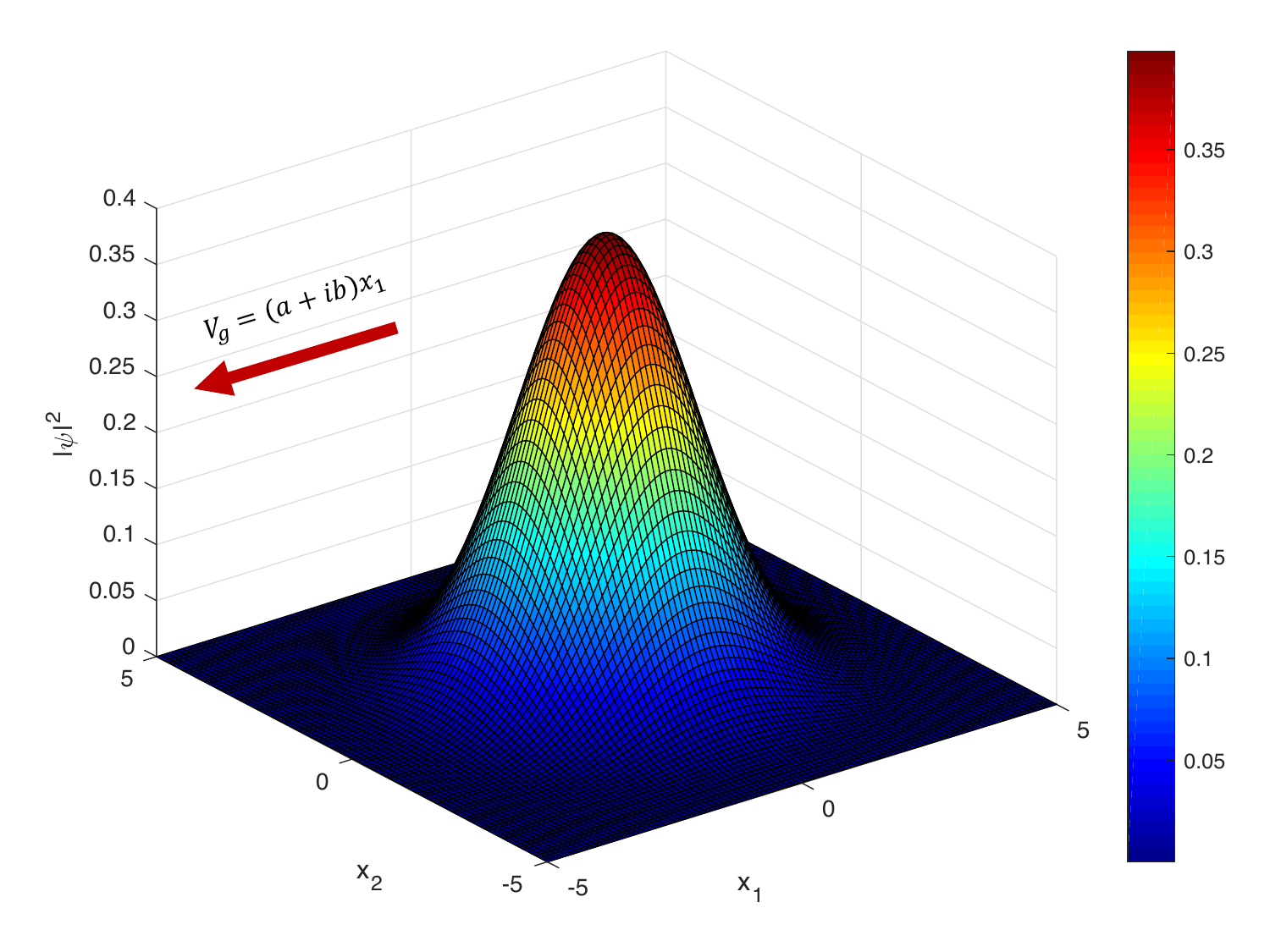}
	\caption{The schematic diagram of BEC in a Non-Hermitian gravitational potential field. Here, it is assumed that the initial BEC wave function satisfies a Gaussian distribution in two-dimensional space. The gradient direction of the Non-Hermitian gravitational potential field, which acts as an external potential field, is parallel to the $x_1$ axis. }
	\label{fig-sketch}
\end{figure}

\emph{Non-Hermitian gravitational effects on BEC.}---%
Consider the BEC with Non-Hermitian gravitational potential (Fig. \ref{fig-sketch}), which characterised by dimensionless GP equation \cite{gpe-01, gpe-02, gpe-03, gpe-04, gpe-05}: 
\begin{equation}
\label{eq-gpe}
	i \frac{\partial}{\partial t} \psi = -\nabla^2 \psi + \eta |\psi|^2 \psi + V_g \psi, 
\end{equation}
where $\psi$ is the wave function of BEC, $t$ is the time, $\nabla^2$ is the Laplace operator, $\eta$ presents the interaction strength, $V_g$ characterises the gravitational potential, all the variables are dimensionless. To investigate the Non-Hermitian effects of the gravitational potential $V_g$, represent it as:
\begin{equation}
\label{eq-vg}
	V_g = (a + i b) x_1,
\end{equation}
where $a$ and $b$ are real valued numbers which characterise the real part and the imaginary part of $V_g$, the $x_1$ characterises one of the dimensionless space coordinates. 


Assume the initial wave function of BEC satisfies a two-dimensional isotropic Gaussian distribution that:
\begin{equation}
\label{eq-initial}
	\psi_0 = \sqrt{ \left(\frac{1}{\sqrt{2 \pi \sigma^2}} \right)^2 e^{-\frac{x_1^2 + x_2^2}{2 \sigma^2}} },
\end{equation}
where the $x_1$ and $x_2$ are the space coordinates, the $\sigma$ is the inflection point of the Gaussian distribution. By solving the GP equation (Eq. (\ref{eq-gpe})) with the initial conditions from Eq. (\ref{eq-initial}), the dynamical evolution of the BEC incorporating a Non-Hermitian gravitational potential can be explored.

\begin{figure*}[ ]
	\centering
	\includegraphics[width = 0.9 \textwidth]{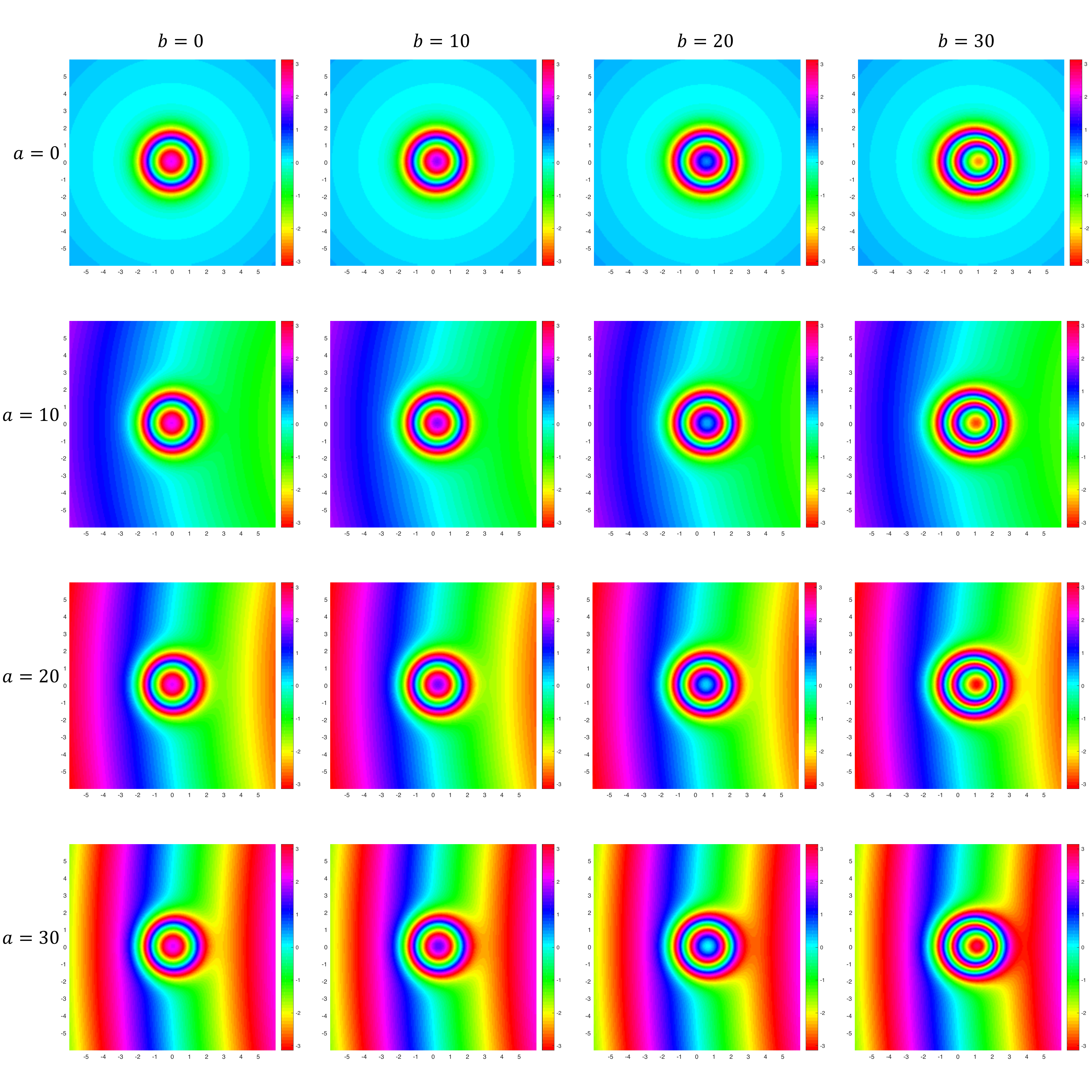}
	\caption{The comparison of the phase distribution of BEC's wave functions under different Non-Hermitian gravitational potential parameters. These results are obtained through numerical solutions of the GP equation. The horizontal and vertical coordinates represent the coordinates of the two spatial dimensions in which the BEC exists, with different colours indicating the values of the wave function's phase. In each row of subplots here, the real part of the gravitational potential energy intensity is the same. The values of $a$ for the 1st, 2nd, 3rd, and 4th rows are $0$, $10$, $20$, and $30$, respectively. Similarly, in each column of subplots, the imaginary part of the gravitational potential energy intensity is the same. The values of $ib$ for the 1st, 2nd, 3rd, and 4th columns are $0i$, $10i$, $20i$, and $30i$, respectively. The parameter $\eta$ representing the interaction strength, is set to $3000$, and the flight time is $0.024$. It can be observed that, as the real part of the gravitational potential strength increases, there is a noticeable change in the phase distribution outside the main peak of the wave function, forming distinct hierarchical structures. Meanwhile, the phase distribution inside the main peak of the wave function remains relatively unchanged. As the imaginary part of the gravitational potential strength increases, the period of the phase distribution inside the main peak of the wave function shortens, and the main peak of the wave function shifts towards regions with stronger gravitational potential. However, the phase distribution outside the main peak of the wave function remains relatively unchanged. }
	\label{fig-phase}
\end{figure*}

The Fig. \ref{fig-phase} illustrates a comparative representation of the phase distribution of a BEC obtained through numerical solutions of the GP equation based on different gravitational parameters. In order to explore the impact of Non-Hermitian gravitational potentials on the BEC wave function, we adjusted the real and imaginary parts of the gravitational parameters separately and performed numerical solutions of the 2D GP equation with these adjusted parameters to obtain the results depicted in the figure.

We know that the wave function of the BEC can be represented as: $\psi = A \exp (i \theta)$, where different colours in the figure represent the values of the wave function phase, $\theta$. Observing the results in the figure, it becomes evident that the real and imaginary parts of the Non-Hermitian gravitational potential have distinct effects on the phase distribution of the wave function.

When the real part of the gravitational potential strength increases, there is a noticeable change in the phase distribution outside the main peak of the BEC wave function, leading to well-defined phase patterns. However, changes in the real part of the gravitational potential strength do not seem to have a significant impact on the phase distribution within the main peak of the BEC wave function.

On the other hand, when the imaginary part of the gravitational potential strength increases, the phase distribution within the main peak of the BEC wave function undergoes a significant change. This change is characterised by a reduction in the spatial period of the wave function phase and a faster variation of the phase with changes in spatial position. Additionally, the peak of the wave function moves towards regions with higher potential energy. Changes in the imaginary part of the gravitational potential strength do not seem to have a significant impact on the phase distribution outside the main peak of the BEC wave function.
We observed the same results in the videos of the dynamical evolution of the BEC, which are included in the Supplemental Material \cite{sm-01}. 
In summary, Non-Hermitian gravitational potentials have a significant influence on the phase distribution of the BEC wave function. Specifically, the impact is observed in terms of the spatial periodicity of the wave function phase and the spatial position of the peak of the wave function, both of which are affected by the imaginary part of the Non-Hermitian gravitational potential strength. 

In order to further explore the impact of Non-Hermitian gravitational potential on the evolution of the BEC wave function, we represent it as follows:
\begin{equation}
	V_g = A_g e^{i \varphi} x_1,
\end{equation}
where $A_g=\sqrt{a^2 + b^2}$ is the magnitude, $\varphi = \arctan (b/a)$ is the complex angle. We maintain a fixed magnitude $A_g$ of the Non-Hermitian gravitational potential and establish a control group by varying the complex angle $\varphi$ in increments of $\pi/8$. We then numerically solve the GP equation, and the obtained results are presented. 

\begin{figure*}
	\centering
	\includegraphics[width = 0.9 \textwidth]{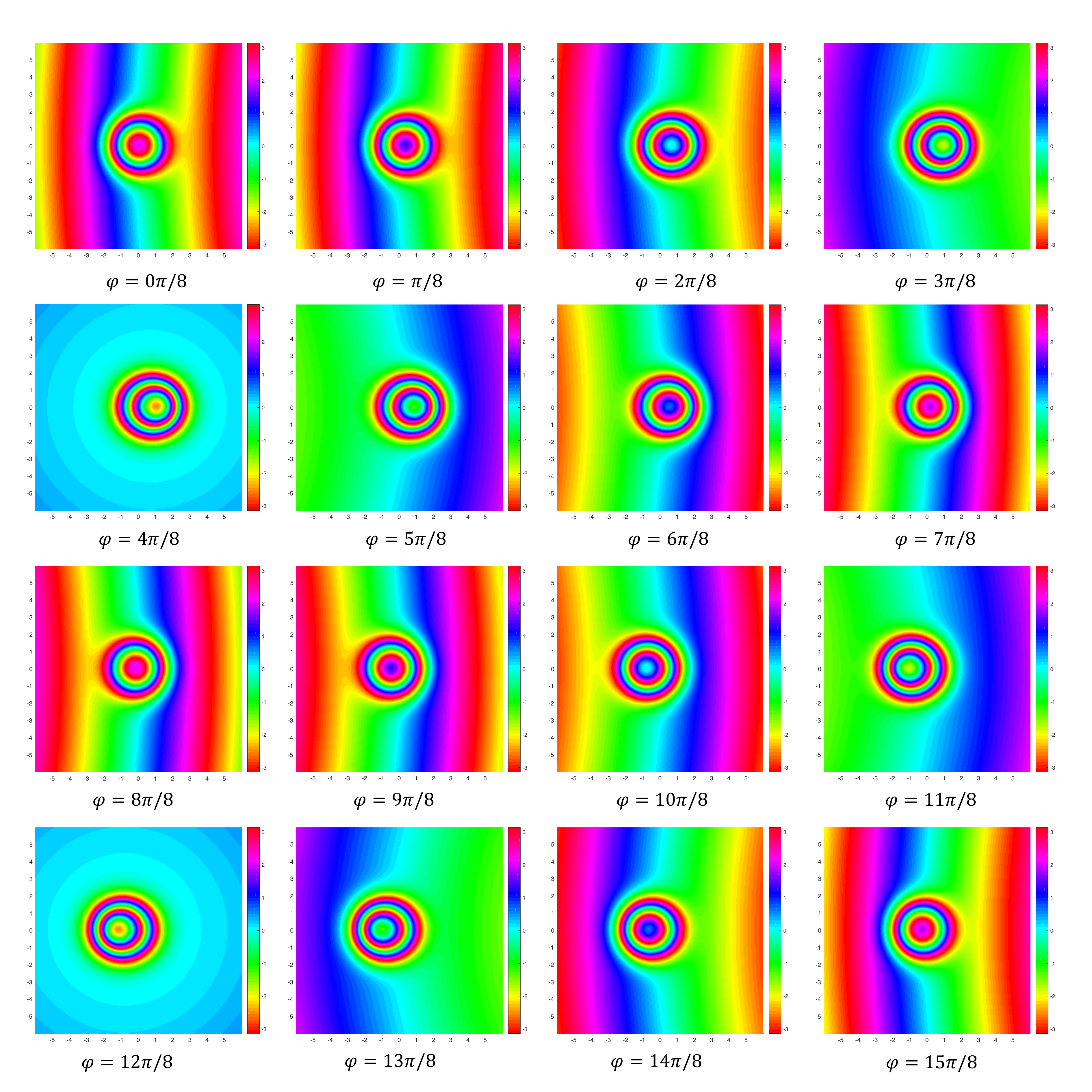}
	\caption{Comparison of phase distributions of BEC wave functions in a two-dimensional space under different Non-Hermitian gravitational potential parameters $\varphi$. These results are obtained from numerical solutions of the GP equation. In the first row, columns 1, 2, 3, and 4 correspond to Non-Hermitian potential complex angles $\varphi$ of $0$, $\pi/8$, $2\pi/8$, and $3\pi/8$, respectively. In the second row, columns 1, 2, 3, and 4 correspond to Non-Hermitian potential complex angles $\varphi$ of $4\pi/8$, $5\pi/8$, $6\pi/8$, and $7\pi/8$, respectively. In the third row, columns 1, 2, 3, and 4 correspond to Non-Hermitian potential complex angles $\varphi$ of $8\pi/8$, $9\pi/8$, $10\pi/8$, and $11\pi/8$, respectively. In the fourth row, columns 1, 2, 3, and 4 correspond to Non-Hermitian potential complex angles $\varphi$ of $12\pi/8$, $13\pi/8$, $14\pi/8$, and $15\pi/8$, respectively. The parameter $\eta$, representing the interaction strength, is set to 3000, the flight time is 0.024, and the amplitude of the gravitational potential $A_g$ is 30. In the same row, adjacent sub-figures have complex angles differing by $\pi/8$, while in the same column, adjacent sub-figures have complex angles differing by $\pi/2$. }
	\label{fig-phase-angle}
\end{figure*}

Observing the changes in the phase distribution of the BEC wave function in Fig. \ref{fig-phase-angle}, one can explore the effects of complex angles on the Non-Hermitian gravitational potential. As the complex angle $\varphi$ increases, the phase distribution of the BEC wave function exhibits periodic variations. In other words, the direct physical quantities influencing the wave function's phase distribution are the real and imaginary parts of the Non-Hermitian gravitational potential. This is consistent with the results observed in Fig. \ref{fig-phase}. It is also noted that when the real part of the Non-Hermitian gravitational potential parameter is greater or less than 0, the phase distribution outside the main peak of the BEC wave function exhibits opposing trends. When the imaginary part of the Non-Hermitian gravitational potential parameter is greater or less than 0, the peak of the BEC wave function's phase distribution exhibits opposite directional shifts. When the complex angles of the Non-Hermitian gravitational potential parameters differ by $\pi$, the phase distribution of the BEC wave function will exhibit mirror symmetry.

Let $a_{mn}$ represent the sub-figures in the $m$-th row and $n$-th column in Fig. \ref{fig-phase-angle}. Observing $a_{14}$ with $a_{22}$, $a_{13}$ with $a_{23}$, $a_{12}$ with $a_{24}$, $a_{11}$ with $a_{31}$, $a_{44}$ with $a_{32}$, $a_{43}$ with $a_{33}$, $a_{42}$ with $a_{34}$, one can see opposing trends in the phase distribution outside the main peak for the real parts of the gravitational parameters. Observing $a_{12}$ with $a_{44}$, $a_{13}$ with $a_{43}$, $a_{14}$ with $a_{42}$, $a_{21}$ with $a_{41}$, $a_{22}$ with $a_{34}$, $a_{23}$ with $a_{33}$, $a_{24}$ with $a_{32}$, one can observe opposite directional shifts in the peak of the phase distribution for the imaginary parts of the opposing gravitational parameters. Notably, the comparison between $a_{21}$ and $a_{41}$ is the most pronounced. Observing $a_{11}$ with $a_{31}$, $a_{12}$ with $a_{32}$, $a_{13}$ with $a_{33}$, $a_{14}$ with $a_{34}$, $a_{21}$ with $a_{41}$, $a_{22}$ with $a_{42}$, $a_{23}$ with $a_{43}$, $a_{24}$ with $a_{44}$, one can observe the mirrored symmetry of the BEC wave function's phase distribution when the complex angles of the gravitational potential parameters differ by $\pi$. 


\emph{Momentum equation for Non-Hermitian gravitational parameters.}---%
To explore the physical essence of these results, we return to the most fundamental form of the dimensionless wave function:
\begin{equation}
	\psi = A e^{i S} = A e^{ i px} e^{-i Ht},
\end{equation}
where $A$ is the amplitude of the wave function, $S=\int L dt$ is the action, $p$ is the momentum, $x$ is the space coordinate, $H$ is the Hamiltonian representing the total energy, $t$ is the time. It can be observed that the spatial frequency of the wave function is characterized by momentum $p$, while the temporal frequency is characterized by total energy $H$. We know that, based on the Hamiltonian canonical equation that: $dp/dt = -\partial H/ \partial x$, the momentum can be represented as: $p = p_0 - \int( \partial H / \partial x )dt$,  where $p_0$ is the initial momentum. Then, we return to the Hamiltonian of the BEC system in Eq. (\ref{eq-gpe}), and substitute it with the momentum operator relation $p= -i \nabla $, thus obtaining: 
\begin{equation}
\label{eq-h}
	H = p^2 + \eta |\psi|^2 + V_g,
\end{equation}
where $p$ represents the momentum of the system at a certain space-time point. 
Let $V_g = (a+ib)x$, and substitute it into the Hamiltonian, 
we can obtain:
\begin{equation}
\label{eq-momentum}
	\frac{d p}{d t} = -2 p \frac{\partial p}{\partial x} - 2\eta |\psi|^2 b t - (a+ib).
\end{equation}
For simplicity, let's assume that the variables in the equation are smooth, and integrate both sides of the equation with respect to time $t$ over a small time interval, thereby obtaining:
\begin{equation}
\label{eq-p-first}
	p = p_0 - 2p \frac{\partial p}{\partial x} t - \eta |\psi|^2 b t^2 -(a+ib)t,
\end{equation}
which is a first-order ordinary differential equation for the function $p$ with respect to the variable $x$. Solving this equation and substituting the initial condition $p(x=0)=p_0$, results in:
\begin{equation}
\label{eq-p-abxt}
\begin{split}
	p = &(\eta |\psi|^2 b t^2 + (a+ib)t ) e^{-\frac{\frac{x}{2t}+p-p_0}{p_0-\eta |\psi|^2 b t^2 - (a+ib)t} } \\
	&+ p_0-\eta |\psi|^2 b t^2 - (a+ib)t,
\end{split}
\end{equation}
where $x$ plays a pivotal role in the variation of $p$. Expanding the exponential term to first order, we can obtain the following expression:
\begin{equation}
	p = p_0 - \frac{x}{2p_0}(\eta |\psi|^2 bt + a + ib  ), 
\end{equation}
which demonstrates the influence of Non-Hermitian gravitational potential parameters on the momentum distribution. 
We know that the real part of the momentum of the wave function corresponds to the de Broglie wavelength. The larger the absolute value of the real part of the momentum, the shorter the de Broglie wavelength, and the smaller the spatial periodicity of the wave function's phase. 
In Eq. (\ref{eq-p-abxt}), it can be observed that at locations with a higher density $|\psi|^2$ of BEC particles, the variation of the imaginary part $b$ of the Non-Hermitian potential parameter can more significantly alter the distribution of the wave function's phase. This is consistent with the variation in the phase distribution within the main peak of the wave function, as shown in Fig. \ref{fig-phase}. 
When $|\psi|^2$ approaches zero, the change in momentum $p$ is determined by the real part $a$ of the Non-Hermitian gravitational potential parameter. In other words, the variation of the real part $a$ of the Non-Hermitian gravitational potential parameter will significantly affect the phase distribution outside the main peak of the BEC wave function. This is also consistent with the results shown in Fig. \ref{fig-phase}. 

 
When considering the scenario with $p_0=0$, solving Eq. (\ref{eq-p-first}) under this condition results in: 
\begin{equation}
	p = (\eta |\psi|^2 b t^2 + (a+ib)t ) \left(e^{\frac{\frac{1}{2t}x + p}{\eta |\psi|^2 b t^2 + (a+ib)t} } -1 \right),
\end{equation}
then, we consider the second-order expansion of the exponential part, thus obtaining:
\begin{equation}
	p^2 + \frac{x}{t}p + \frac{x^2}{4t^2} + \frac{(\eta |\psi|^2 b t^2 + (a+ib)t)x}{t} = 0,
\end{equation}
which can be considered as a quadratic equation in terms of momentum $p$, and solving this equation yields: 
\begin{equation}
	p = -\frac{x}{2t} \pm \sqrt{-\eta |\psi|^2 b t - (a+ib) } \sqrt{x},
\end{equation}
which can be approximated by:
\begin{equation}
	p = -\frac{x}{2t} \pm \left(-\frac{1}{2} \frac{b}{\sqrt{\eta |\psi|^2 b t + a} } + i \sqrt{\eta |\psi|^2 b t + a} \right) \sqrt{x}, 
\end{equation}
where the $(\eta |\psi|^2 b t + a)$ is supposed to be greater than than $b$. Based on this expression, we can observe that as time $t$ gradually increases, the real part of momentum $p$ will decrease, while the imaginary part of momentum $p$ will increase. This will lead to a faster exponential decay of the wave function's amplitude, resulting in an exponential decay of the BEC particle density, causing a spatial shift of the peak of the wave function.
The symbol $\pm$ here is used to illustrate the symmetry in the wave function's phase distribution in Fig. \ref{fig-phase-angle} concerning the complex angles of the Non-Hermitian gravitational potential parameter.

Conservation laws play a crucial role in describing preserved physical quantities within nonlinear evolution equations (NLEEs) in mathematics \cite{math-01, math-02}. These laws serve as integral characteristics for NLEEs, allowing the establishment of uniqueness, existence, and stability analysis of solutions \cite{math-03}. By leveraging these conserved quantities, we can also derive conservation laws and identities for the equation and its derivatives through intricate calculations. Moreover, let $p=p(x,t)$ represent the smooth solution of the momentum equation Eq. (\ref{eq-momentum}) with an initial value of  $p(x,0)=p_{0}(x).$ We obtain the conservation law:
\begin{equation}
	\frac{\partial}{\partial t} \left( p(x, t)+\eta|\psi|^{2}bt^{2}+(a+ib)t \right)+\frac{\partial}{\partial x} \left( |p(x, t)|^{2} \right) = 0. 
\end{equation}
Considering the boundary conditions $p(x, t)|_{x\rightarrow\pm\infty}\rightarrow 0,$ we get: 
\begin{equation}
	\frac{d}{dt}\int_{\mathbb{R}} \left( p(x, t)+\eta|\psi|^{2}bt^{2}+(a+ib)t \right) dx=0. 
\end{equation}
Then in this context, a conservation law can be expressed as follows: 
\begin{equation}
	E(\cdot, t)=E_{0}, 
\end{equation}
with:
\begin{equation}
\begin{split}
	&E(\cdot, t) = \int_{\mathbb{R}} \left(p(x, t)+\eta|\psi|^{2}bt^{2}+(a+ib)t \right)dx,\\ 
	&E_{0} = \int_{\mathbb{R}}p_{0}(x)dx. 
\end{split}
\end{equation}

It can be observed that within a certain range, $p(x, t)+\eta|\psi|^{2}bt^{2}+(a+ib)t$ is a conserved quantity. For the case of $p_0=0$, when the imaginary part $b$ of the Non-Hermitian parameter increases, due to the effect of $\eta|\psi|^{2}bt^{2}$, in the region where $|\psi|^2 \gg 1/(\eta t)$, the absolute value of the real part of $p$ will increase, which will cause the spatial period of the phase of the wave function from the center to the outside of the wave function main peak to decrease. At the same time, due to the effect of $(a+ib)t$, the absolute value of the imaginary part of $p$ will increase, which will cause the exponential decay of the wave function amplitude with respect to space, manifested as the spatial displacement of the wave function peak. When the real part $a$ of the Non-Hermitian parameter increases, due to the effect of $(a+ib)t$, the absolute value of the real part of $p$ will increase, which will cause the spatial period of the wave function along the horizontal axis to decrease. These observations are consistent with our numerical results (Fig. \ref{fig-phase}, Fig. \ref{fig-phase-angle}, and videos in the Supplementary Material \cite{sm-01}).



\emph{Conclusion.}---%
We investigated the impact of Non-Hermitian gravitational potentials on the phase distribution of BEC wave functions. We observed that an increase in the imaginary part of Non-Hermitian gravitational potential parameters leads to a reduction in the spatial periodicity of the BEC wave function phase, accompanied by spatial displacement of the peak. Constructing equations describing the relationship between the momentum of the BEC wave function and Non-Hermitian gravitational potential parameters, we explained the variation in the wave function spatial periodicity, characterized by the de Broglie wavelength associated with momentum, through solving this equation under specific conditions. This explanation aligns with our numerical results.
Our research contributes to exploring the physical essence of Non-Hermitian gravitational potentials and their impact on BEC. Additionally, it provides theoretical guidance for related BEC experiments in this field.


\emph{Acknowledgments.}---%
This work was supported by National Key R\&D Program of China under grants No. 2021YFA1400900, 2021YFA0718300, 2021YFA1402100, NSFC under grants Nos. 12174461, 12234012, 12334012, 52327808, Space Application System of China Manned Space Program.



\bibliography{CITE.bib}

\end{document}